\documentclass[11pt]{article}
\usepackage{moriond,epsfig}

\def\Journal#1#2#3#4{{#1} {\bf #2}, #3 (#4)}

\def\NIM{\em Nucl. Instrum. Methods}

\def\NPBs{{\em Nucl. Phys.} B (Proc. Suppl.)}
\def\PLB{{\em Phys. Lett.}  B}
\def\PRL{\em Phys. Rev. Lett.}
\def\PRD{{\em Phys. Rev.} D}

\def\APJ{\em Astrophysical Jornal}

\def\be{\begin{equation}}
\def\ee{\end{equation}}
\def\bea{\begin{eqnarray}}
\def\eea{\end{eqnarray}}

\begin{document}
\vspace*{4cm}
\title{Super Kamiokande results: atmospheric and solar neutrinos}

\author{M.Ishitsuka\\
        for the Super-Kamiokande collaboration }

\address{Institute for Cosmic Ray Research, University of Tokyo \\ 
        5-1-5 Kashiwanoha, Kashiwa 277-8582, Japan}

\maketitle\abstracts{Atmospheric neutrino and solar neutrino data from
 the first phase of Super-Kamiokande (SK-I) are presented.
  The observed data are used to study atmospheric and solar
 neutrino oscillations. Zenith angle distributions from various atmospheric
 neutrino data samples are used to estimate the neutrino oscillation
 parameter region. In addition, a new result of the $L/E$ measurement
 is presented. A dip in the $L/E$ distribution was observed
 in the data, as predicted from the sinusoidal flavor transition
 probability of neutrino oscillation.  The energy spectrum and the time
 variation such as day/night and seasonal differences of solar neutrino
 flux are measured in Super-Kamiokande. The neutrino oscillation parameters
 are strongly constrained from those measurements.}

\section{Introduction}
 Super-Kamiokande is a 50,000\,ton water Cherenkov detector located 
1,000\,m (2,700\,m water equivalent) under Mt. Ikenoyama at Kamioka Observatory,
 Gifu Prefecture, Japan. The detector is a cylindrical tank and is optically divided
 into two regions. The inner detector (ID) is instrumented with 11,146 inward facing
 20\,inch PMTs which give a photo cathode coverage of 40\,\%. The outer detector (OD)
 completely surrounds the ID with the thickness of 2.05\,m to 2.2\,m water and is monitored
 by 1,885 outward-facing 8\,inch PMTs.  The OD works as a veto counter 
against cosmic ray muons. The charge information observed in the OD is also used to
separate event sample in the $L/E$ analysis.

\section{Atmospheric neutrinos}

\subsection{Introduction}

Atmospheric neutrinos were observed in Super-Kamiokande during a 1489\,live day
exposure which corresponds to 92\,kiloton-yr. The atmospheric neutrino events are
classified into fully contained (FC), partially contained (PC) and upward going muons.
 The vertices of neutrino interactions are required to be inside the fiducial
 volume of the ID for FC and PC events.
If the tracks of entire particles are contained inside the ID, the event is classified
into FC. While one of the particles, mostly muon, exits the ID and deposits visible
energy in the OD for PC events.  Each observed Cherenkov ring is identified as
either $e$-like or $\mu$-like based on the ring pattern.
 The directions and the momentum of charged particles
can be reconstructed from the ring image.
Upward going muons are also observed as an atmospheric neutrino sample. These muons
originate from high energy neutrino interactions with the rock surrounding the detector.
Upward going muon events are classified into upward stopping muons having only an
 entrance signal in the OD, and upward through-going muons having both entrance and
 exit signals.
 The atmospheric neutrino events in Super-Kamiokande are predicted by a detailed Monte
Carlo simulation~\cite{atmpd}. 

\subsection{Zenith angle analysis}\label{subsec:zenith}

\begin{figure}[htb]
\begin{center}
  \includegraphics[height=2.5in]{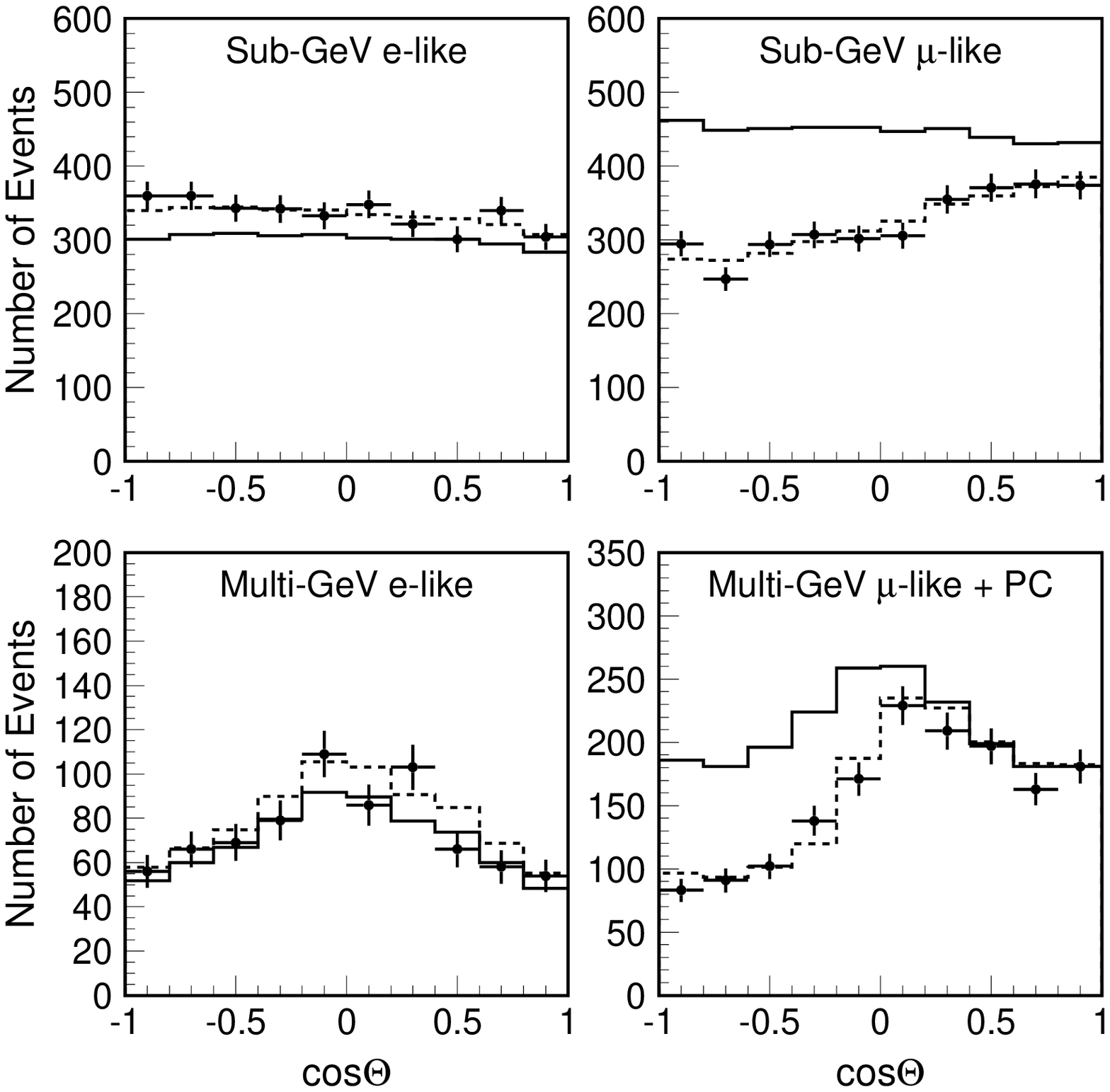} 
  \includegraphics[height=2.5in]{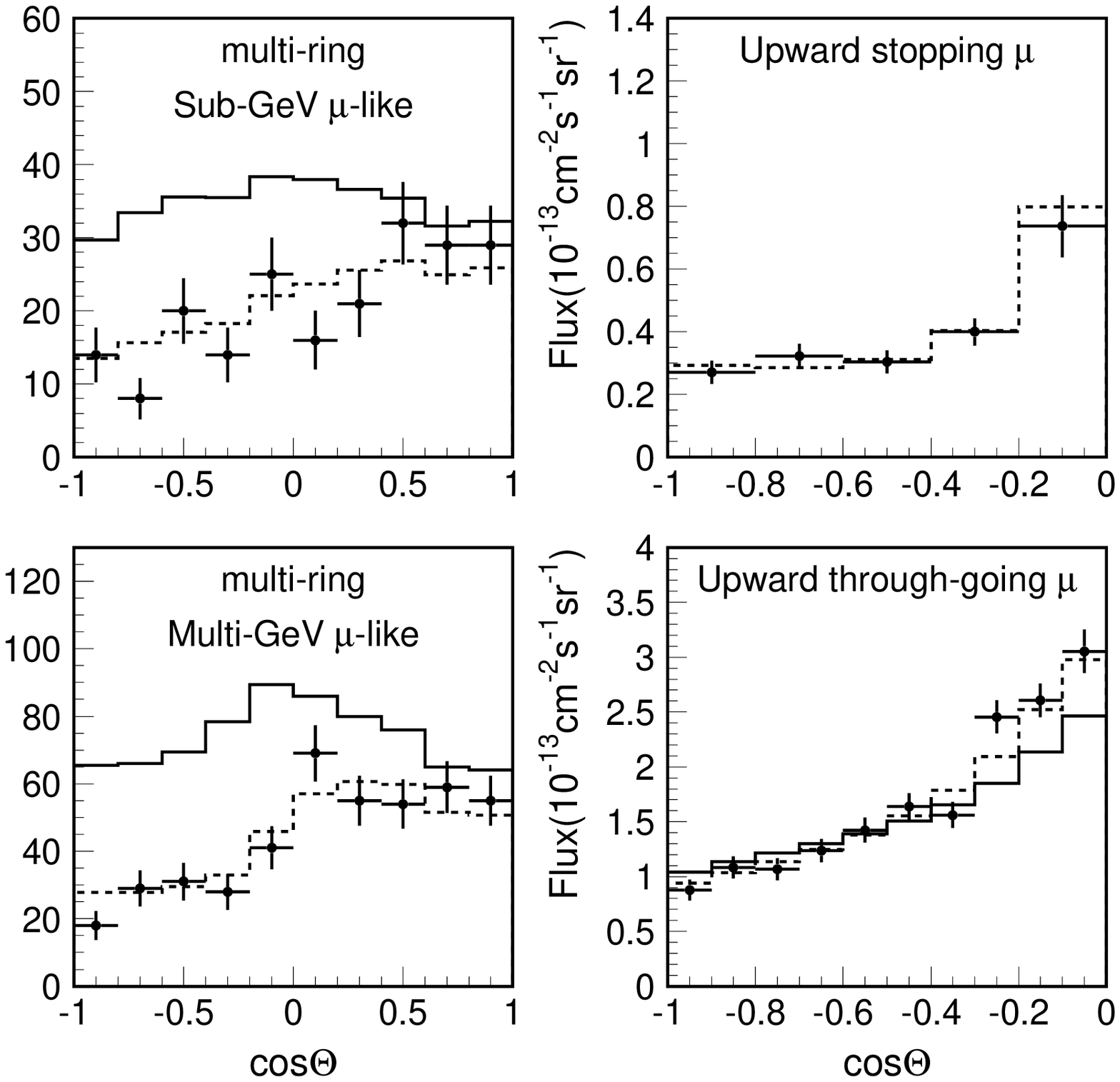} 
  \caption{Zenith angle distribution for fully-contained single-ring $e$-like and $\mu$-like events, multi-ring 
$\mu$-like events, partially contained events and upward-going muons. The points show the data and the solid
lines show the Monte Carlo events without neutrino oscillation.
 The dashed lines show the best-fit expectations for $\nu_\mu \leftrightarrow \nu_\tau$
 oscillations.}
 \label{fig:zenith_angle}
\end{center}
\end{figure}

Figure~\ref{fig:zenith_angle} shows the zenith angle distributions of each atmospheric neutrino sample.
The data are compared with the MC expectation.
 The observed muon neutrino events exhibit a strong
zenith angle dependent deficit compared with the expectation without neutrino oscillation.
On the other hand, the observed electron neutrino events are consistent with the prediction within about 10\,\% normalization.

A fit to the FC, PC and upward-going muon data is carried out
assuming 2-flavor $\nu_\mu \leftrightarrow \nu_\tau$ oscillation.
In 2-flavor $\nu_\mu \leftrightarrow \nu_\tau$ oscillations, survival probability 
of $\nu_{\mu}$ is expressed by~: 
\begin{equation}
\label{eqn:oscillation}
 P(\nu_{\mu} \rightarrow \nu_{\mu})
= 1-\sin^22\theta\sin^2\left(\frac{1.27\Delta{m}^2({\rm eV}^2) L({\rm km})}{E({\rm GeV})}\right),
\end{equation}
where $E$ is the neutrino energy and $L$ is the flight length of neutrinos.
 Since the observed
zenith angle distributions of $e$-like events agree with the predictions,
2-flavor $\nu_\mu \leftrightarrow \nu_\tau$ oscillation is considered to be dominant 
in the atmospheric neutrino oscillations. A $\chi^2$ test is employed to evaluate
the agreement of the fit to the observed data. The FC and PC samples are divided into 10 bins
equally spaced between $\cos \Theta =-1$ and $\cos \Theta = +1$, where $\Theta$ is 
zenith angle of particle direction. Furthermore, the FC sample is divided by the momentum,
number of rings (single-ring and multi-ring) and particle types (e-like and $\mu$-like).
The upward stopping and upward through-going muon samples are divided into 5 and 10 bins
equally spaced between $\cos\Theta=-1$ and $\cos \Theta = 0$, respectively.
In total 175 bins are used in the zenith angle analysis.

\begin{figure}[htb]
\begin{center}
  \includegraphics[height=2.1in]{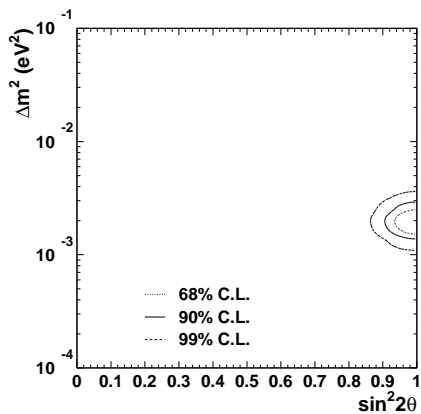} 
  \caption{Allowed oscillation parameter regions for $\nu_{\mu} \leftrightarrow \nu_{\tau}$ oscillations.
Three contours correspond to the 68\,\% (dotted line), 90\,\% (solid line) and 99\,\% 
(dashed line) C.L. allowed regions, respectively.}
 \label{fig:zenith_allowed}
\end{center}
\end{figure}

A global scan was carried out on a $(\sin^22\theta, \Delta m^2)$ grid including
 the unphysical region ($\sin^22\theta > 1$). The best-fit for 2-flavor 
$\nu_\mu \leftrightarrow \nu_\tau$ oscillations was obtained at
$(\sin^22\theta = 1.0, \Delta m^2 = 2.0\times10^{-3} $~eV$^2$) in which $\chi^2_{min} = 170.8 / 172 {\rm ~DOF}$.
Figure~\ref{fig:zenith_allowed} shows the contour plot of the allowed 
neutrino oscillation parameter regions. Three contours
 correspond to the 68\,\%, 90\,\% and 99\,\% confidence levels (C.L.), respectively.
The allowed oscillation parameter regions were 
$\sin ^2 2 \theta > 0.90$ and
$1.3 \times 10^{-3} < \Delta m^2 < 3.0 \times 10^{-3} $ eV$^2$
at 90\,\%~C.L.

\subsection{$L/E$ analysis}

Zenith angle dependent deficit of muon neutrinos has been interpreted as evidence for 
atmospheric neutrino oscillations~\cite{atmpd98}.
Neutrino oscillations have been studied using various neutrino sources.
  However, the sinusoidal neutrino flavor transition probability predicted by neutrino
oscillation has not been demonstrated yet.
 The analysis described herein used a selected
sample of atmospheric neutrino events, those with good resolution in
$L/E$, to search for the dip in oscillation probability expected when the
argument of the second sine-squared term in Eq.~\ref{eqn:oscillation} is $\pi/2$.

In the $L/E$ analysis, 1489\,live-days exposure of FC $\mu$-like and PC atmospheric neutrino data are used.
Event selection and classification in $L/E$ analysis are different from those in the
zenith angle analysis. In order to increase the statistics of the data, especially of high energy muons,
 the fiducial volume for the FC sample is expanded from 22.5\,kton to 26.4\,kton.
 Estimated non-neutrino background events in the
expanded fiducial volume is less than 0.1\,\% and is negligibly small.
 The PC events is subdivided into two categories~:
 ``OD stopping events'' where the muon stops in the outer detector, and
``OD through-going events'' where the muon exit into the rock.
 The division is based on the amount of Cherenkov
 light detected in the OD.
 Since these two samples have different resolution in $L/E$,
 different cuts were applied for each sample,
 improving the overall efficiency.

The neutrino energy is estimated from the total energy of charged
 particles observed in the ID. The energy deposited in the OD is
 estimated from the potential track length in the OD and is taken into 
 account for PC events. The relationship between the neutrino
 energy and the observed energy is determined based on
 the Monte Carlo simulation.
 The flight length of neutrinos, which ranges from approximately 15\,km to
 13,000\,km depending on the zenith angle, is estimated
 from the direction of the total momentum of the observed particles. 

\begin{figure}[htb]
\begin{center}
  \includegraphics[height=1.65in]{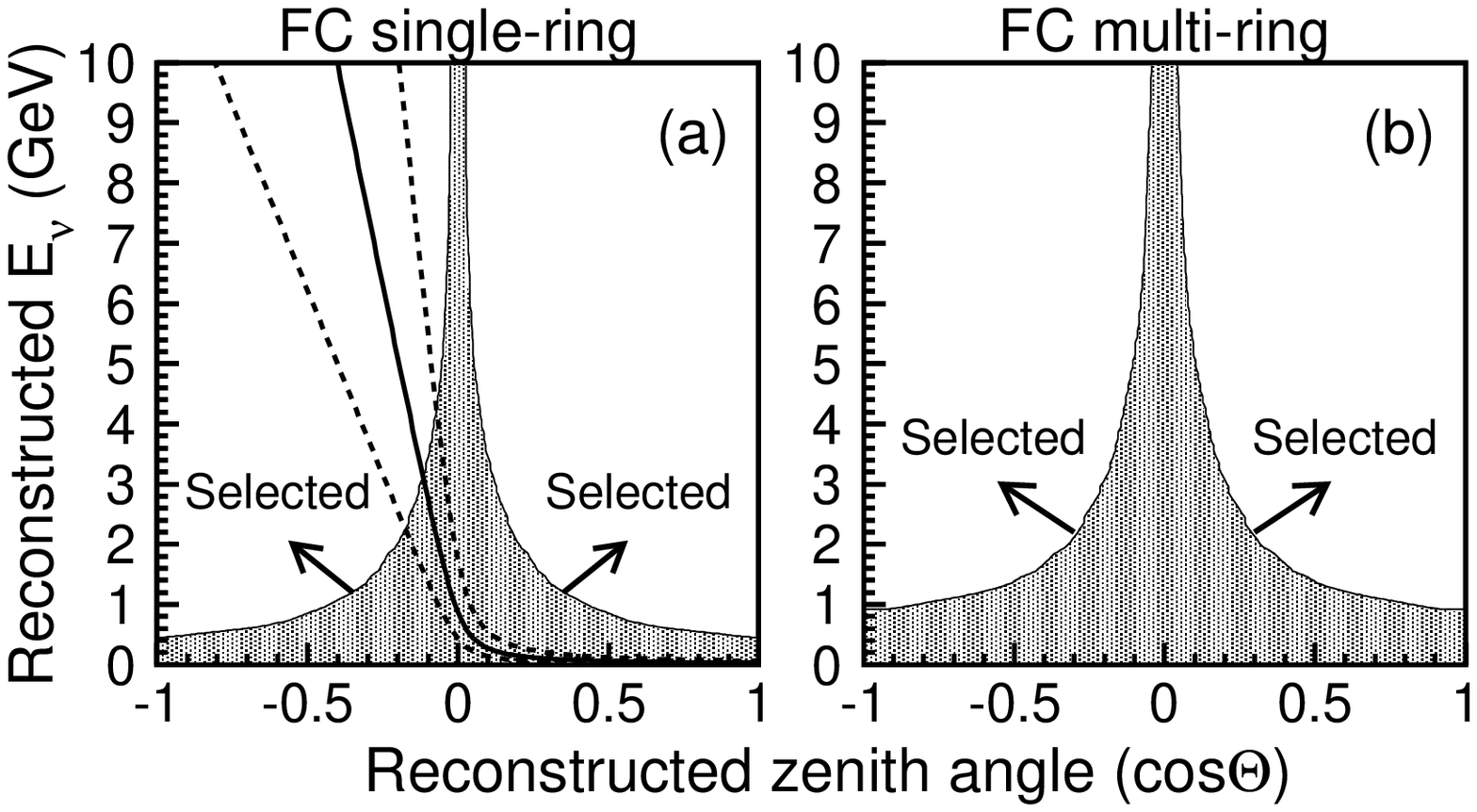}\hspace{0.1in}
  \includegraphics[height=1.65in]{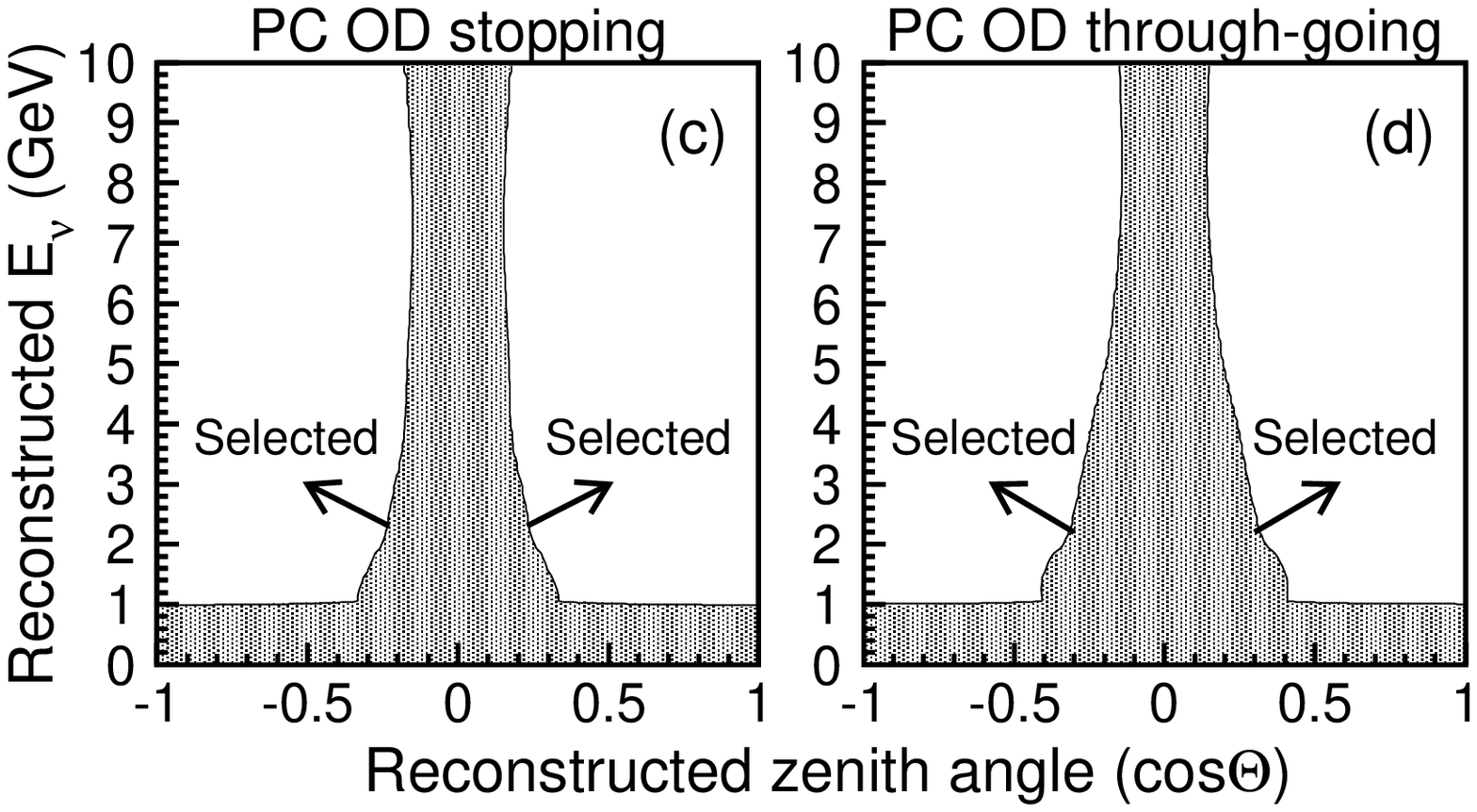} 
  \caption{Contour plots of 70\,\% $L/E$ resolution on the
 ($\cos\Theta,\,\,E_{\nu}$) plane for (a) FC single-ring, (b) FC multi-ring, (c) PC OD stopping
 and (d) PC OD through-going samples. Three lines in the upper
 left figure show the survival probabilities of muon neutrinos predicted
 from neutrino oscillation with ($\sin^2 2\theta,\,\Delta m^2) = (1.00,\,2.4\times10^{-3} $\,eV$^2$).
 Full and half oscillation occur on the solid and dashed lines, respectively. }
 \label{fig:resolution_le}
\end{center}
\end{figure}

The resolution of the reconstructed $L/E$ is calculated at each point on the ($\cos\Theta, E_{\nu}$)
plane, where $\Theta$ is the zenith angle. Figure~\ref{fig:resolution_le} shows 70\,\% $L/E$
resolution contours which are used as the selection criteria. The $L/E$ resolution cut
is set to be $\Delta(L/E) < 70$\,\% from the Monte Carlo simulation to maximize the sensitivities to
distinguish neutrino oscillation from other hypotheses.
Three lines in Figure~\ref{fig:resolution_le} (a) indicate 
survival probabilities of muon neutrinos predicted from neutrino oscillations. Statistics
of high energy muon events is crucial to observe the first oscillation minimum in $L/E$.

\begin{figure}
\begin{center}
  \includegraphics[width=2.6in]{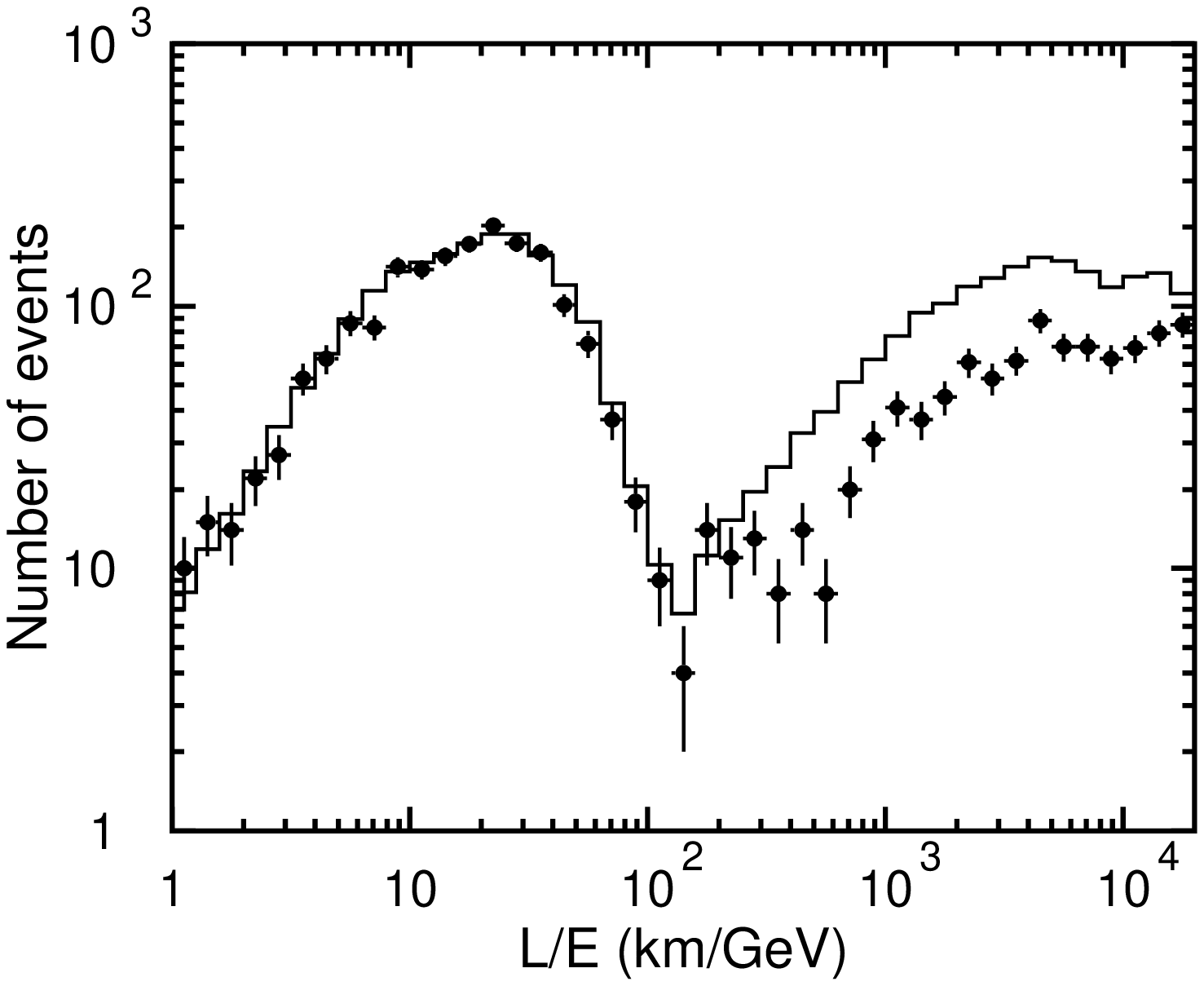} \hspace{0.2in}
  \includegraphics[width=2.6in]{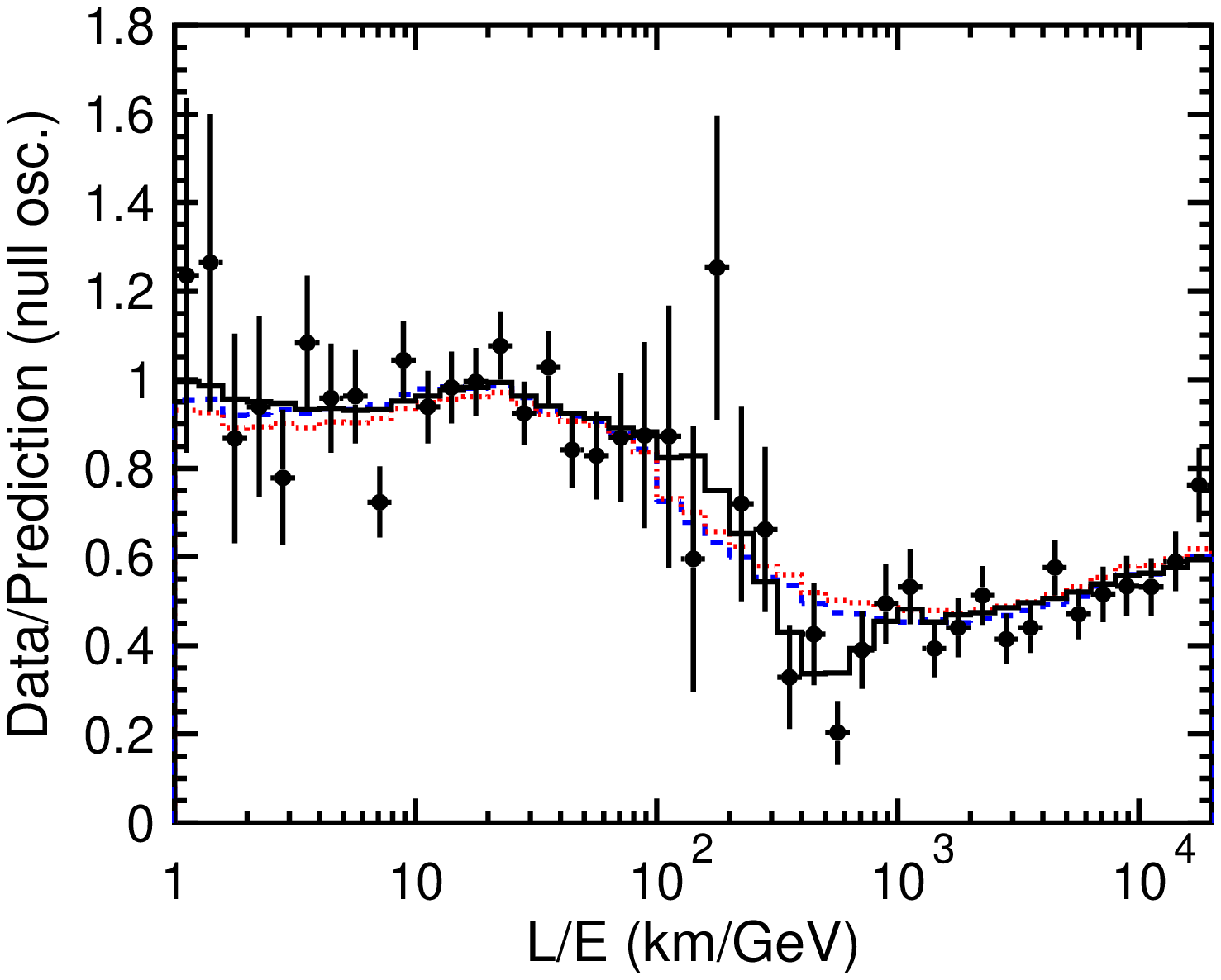}
  \caption{Left: Number of events as a function of the reconstructed $L/E$ 
for the data (points) and the atmospheric neutrino Monte Carlo events (histogram).
 Right: Ratio of the data to the non-oscillated Monte Carlo events (points)
 with the best-fit expectation
 for 2-flavor $\nu_\mu \leftrightarrow \nu_\tau$ oscillations (solid line).
 Also shown are the best-fit expectation for
 neutrino decay (dashed line) and neutrino decoherence (dotted line).}
   \label{fig:le_plot}
\end{center}
\end{figure}

The left-hand plot in Figure~\ref{fig:le_plot} shows the number of events as a function
 of $L/E$ for the data and Monte Carlo predictions, and the right-hand plot shows the data
over non-oscillated Monte Carlo ratio with the best-fit expectation 
for 2-flavor $\nu_\mu \leftrightarrow \nu_\tau$ oscillations in which systematic errors are considered. 
A dip, which should correspond to the first oscillation minimum, was observed around $L/E$~=~500~km/GeV.

A fit to the observed $L/E$ distribution was carried out assuming neutrino
 oscillations. In the analysis, the $L/E$ distribution is divided into
 43 bins from $log(L/E)=0.0$ to 4.3\,.
 The likelihood of the fit and the $\chi^2$ are defined as~:
\begin{eqnarray}
{\cal L}(N^{\rm prd},N^{\rm obs}) = \prod_{i=1}^{43}\frac{\exp{(-N_{i}^{\rm prd})(N_{i}^{\rm prd})^{N_{i}^{\rm obs}}}}{N_{i}^{\rm obs}!} \times\prod_{j=1}^{24}\exp{\left( - \frac{\epsilon_j^2}{2\sigma_j^2} \right)}, \\
\chi^2 \equiv -2\ln\left(\frac{{\cal L}(N^{\rm prd},N^{\rm obs})}{{\cal L}(N^{\rm obs},N^{\rm obs})}\right),~~~~~~~~~~
\label{equation:chi2def_le}
\end{eqnarray}
where $N^{\rm obs}_i$ is the number of the observed events in the $i$-th bin and $N^{\rm prd}_i$
is the number of predicted events, in which neutrino oscillation and systematic
 uncertainties are considered. 25 systematic uncertainties are considered in the $L/E$ analysis,
which include uncertainty parameters from the neutrino flux calculation, neutrino interaction models
and detector performance. Among these, only 24 constrain the likelihood as the absolute normalization 
is allowed to be free. The second term in the likelihood definition represents the contributions
from the systematic errors, where $\sigma_j$ is the estimated uncertainty in the parameter $\epsilon_j$.

\begin{figure}
\begin{center}
  \includegraphics[width=2.3in]{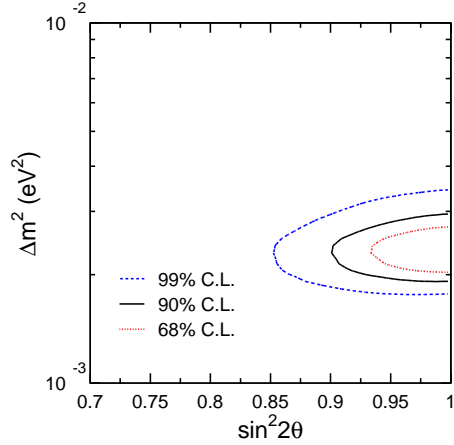}
  \caption{68, 90 and 99\,\% C.L. allowed oscillation parameter regions for
 2-flavor $\nu_\mu \leftrightarrow \nu_\tau$ oscillations obtained
 by the $L/E$ analysis.}
   \label{fig:osc_allowed_le}
\end{center}
\end{figure}

A scan was carried out on a $(\sin^22\theta,\,\log \Delta m^2)$ grid, minimizing $\chi^2$ 
 by optimizing the systematic error parameters. The $\chi^2_{min}$ value was $37.9/40$\,DOF obtained at
 $(\sin^{2}2\theta,\,\Delta m^2)=(1.00,\,2.4\times10^{-3} $\,eV$^2$).
 Including unphysical parameter region ($\sin^22\theta > 1$), the best-fit was obtained at
 $(\sin^{2}2\theta,\,\Delta m^2)=(1.02,\,2.4\times10^{-3} $\,eV$^2$), in which the minimum $\chi^2$
was 0.12 lower than that in the physical region. Figure~\ref{fig:osc_allowed_le} shows the contour plot of the 68, 90 and 99\,\% C.L. allowed oscillation parameter regions.
The 90\,\% C.L. allowed parameter region is obtained as 
 $1.9 \times 10^{-3}\,{\rm eV}^2 < \Delta m^2 
< 3.0 \times 10^{-3}\,{\rm eV}^2 $ and 
$\sin^{2}2\theta~ > 0.90$.
The location of the allowed region is consistent with that of the zenith angle analysis.

The observed $L/E$ distribution was also fit assuming neutrino decay~\cite{decay1,decay2} and neutrino 
decoherence~\cite{decoh1,decoh2}. The $\nu_\mu$ survival probability for neutrino decay 
is expressed as 
$P(\nu_{\mu} \rightarrow \nu_{\mu})$ = $\left[\sin^{2}\theta+\cos^{2}\theta \exp\left(-m/2\tau \cdot L/E\right)\right]^2$ 
where $\tau$ is the lifetime of a neutrino mass state, and that for neutrino decoherence is expressed as
$P(\nu_{\mu} \rightarrow \nu_{\mu})$ = $1-\frac{1}{2}\sin^{2}2\theta\left[1-\exp\left(-\gamma_{0} L/E\right)\right]$ 
where $\gamma_0$ is the decoherence parameter. 
The right-hand plot in Figure~\ref{fig:le_plot} includes 
the best-fit expectation for neutrino decay and decoherence.
The $\chi^2_{min}$ values were $49.1/40$\,DOF
 at $(\cos^{2}\theta,\,m/\tau)=(0.33,\,1.26 \times 10^{-2}$\,GeV/km)
 for neutrino decay and $52.4/40$\,DOF
 at $(\sin^{2}2\theta,\,\gamma_{0})=(1.00,\,1.23\times10^{-21}$\,GeV/km)
 for neutrino decoherence. These $\chi^2_{min}$ values are 11.3 (3.4 standard deviations)
 and 14.5 (3.8 standard deviations) larger than that for neutrino
 oscillation. Alternative models that could explain the zenith-angle and
energy dependent deficit of muon neutrinos are disfavored, since they do not predict the dip structure 
in the $L/E$ distribution.

 The observed $L/E$ distribution, especially the first dip, gives the first direct evidence that the neutrino flavor
transition probability obeys the sinusoidal function as predicted by neutrino flavor oscillations. 

\section{Solar neutrinos}

\subsection{Introduction}

The observed solar neutrino flux in all the experiments has been significantly
smaller than that was expected. From the recent solar neutrino data,
especially together with Super-Kamiokande and SNO, the cause was identified to be
neutrino oscillations~\cite{solar}. In order to determine the neutrino oscillation
parameters, it it important to measure not only solar neutrino flux
 but also the energy spectrum and the time variations,
which are independent of the uncertainties in the solar models.

Solar neutrinos are observed in Super-Kamiokande with a large statistics owing
to its large target volume.  The detector can measure the energy spectrum of neutrino
events precisely by the well calibrated performance\cite{linac}.
Since the observation is working in real-time, the time dependence of solar neutrino flux,
 day/night or seasonal differences, can be also measured. 
In this paper, the results from 1496 days of solar neutrino data observed from May in 1996 to
July in 2001 during the first phase of Super-Kamiokande are reported.

\subsection{Results of solar neutrino observations}

The solar neutrino signals observed in Super-Kamiokande are separated from background events
by fitting the forward peak to the solar direction. The observed solar neutrino flux in
 Super-Kamiokande, whose energy threshold is 5.0MeV, is~:
\begin{equation}
  2.35 \pm 0.02 (stat.) \pm 0.08 (sys.) \quad [\times 10^6/cm^2/sec].
\end{equation}
As compared with the result to standard solar model (BP2000)\cite{ssm}, the ratio is~:
\begin{equation}
  \frac{Data}{SSM} = 0.465 \pm 0.005 (stat.) ^{+0.016}_{-0.015} (sys.),
\end{equation}
The observed flux is significantly smaller than the prediction. The day-night flux differences are also observed.
 A simple day/night asymmetry is obtained by dividing the data sample into day and night as follows~:
\begin{equation}
  A_{\mbox{\small DN}}=\frac{\phi_{day}-\phi_{night}}{(\phi_{day}+\phi_{night})/2}
 = -0.021 \pm 0.020 (stat.)  ^{+0.013}_{-0.012} (sys.),
\end{equation}
which is consistent with zero.
\subsection{Solar neutrino oscillation analysis}

\begin{figure}[tb]
\centerline{\includegraphics[width=2.5in]{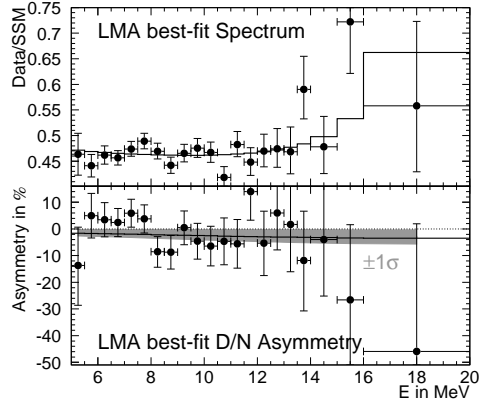}}
\caption{Observed spectrum (top) and D/N asymmetry (bottom).
The predictions (solid lines) are for $\tan^2\theta=0.55$ and
$\Delta m^2=6.3\times10^{-5}$eV$^2$ with 
$\phi_{^8B}=0.96$SSM and
$\phi_{\mbox{\tiny\em hep}}=3.6$SSM.
Each energy bin is fit independently to the rate (top)
and the day/night asymmetry (bottom).
The gray band is the $\pm1\sigma$ range.}
\label{fig:lmadnfit}
\end{figure}

Maximum likelihood fit was carried out to determine the allowed neutrino oscillation parameters considering
the time variations of the solar neutrino flux~\cite{smy}. The data sample is divided
into 21 energy bins from 5\,MeV to 20\,MeV. Two types of probability distribution functions are considered. $p(\cos\theta{\mbox{\tiny sun}},E)$
describes the angular shape expected for solar neutrino signals of energy $E$ and $u_i(\cos\theta{\mbox{\tiny sun}})$
 is the background shape in the energy bin $i$, where $\theta{\mbox{\tiny sun}}$ is the angle between the
 reconstructed recoil electron direction and the solar direction. The likelihood~:
\begin{equation}
{\cal L}=e^{-\left(\sum_i B_i+S\right)}\prod_{i=1}^{N_{\mbox{\tiny bin}}}\prod_{\kappa=1}^{n_i}
\left(B_i\cdot u_i(\cos\theta_{i\kappa})+S\frac{\mbox{MC}_i}{\sum_j\mbox{MC}_j}\cdot p(\cos\theta_{i\kappa},E_\kappa)\cdot z(\alpha,t_\kappa)\right)
\end{equation}
is maximized with the respect to the signal $S$ and the 21 backgrounds $B_i$. MC$_i$ is the number of events expected
in the energy bin $i$. To take into account the time variations of the solar neutrino flux in the likelihood fit,
the signal term is modified by $z(\alpha,t_\kappa)$, where $t_\kappa$ is the event time and $\alpha$ is an
amplitude scaling factor. Solar zenith angle (day/night) variations and the additional seasonal variation are considered.
 For the best-fit LMA parameters, the amplitude scaling factor is obtained as $\alpha=0.86\pm0.77$, which corresponds
 to the day/night asymmetry~:
\begin{equation}
A_{\mbox{\small DN}}=-0.018\pm0.016\mbox{(stat.)}^{+0.013}_{-0.012}\mbox{(sys.)}
\end{equation}
where $-2.1\,\%$ is expected from the simple day/night comparison. The statistical uncertainty is reduced by 25\,\% with this likelihood analysis.
However, the resulting day/night asymmetry is still consistent with zero.
Figure~\ref{fig:lmadnfit} shows
the resulting fitted rate (top) and the day/night asymmetry (bottom)
for each energy bin with the expectations for the best-fit LMA oscillation parameters.

\begin{figure*}[tb]
\begin{center}
\includegraphics[height=2.45in]{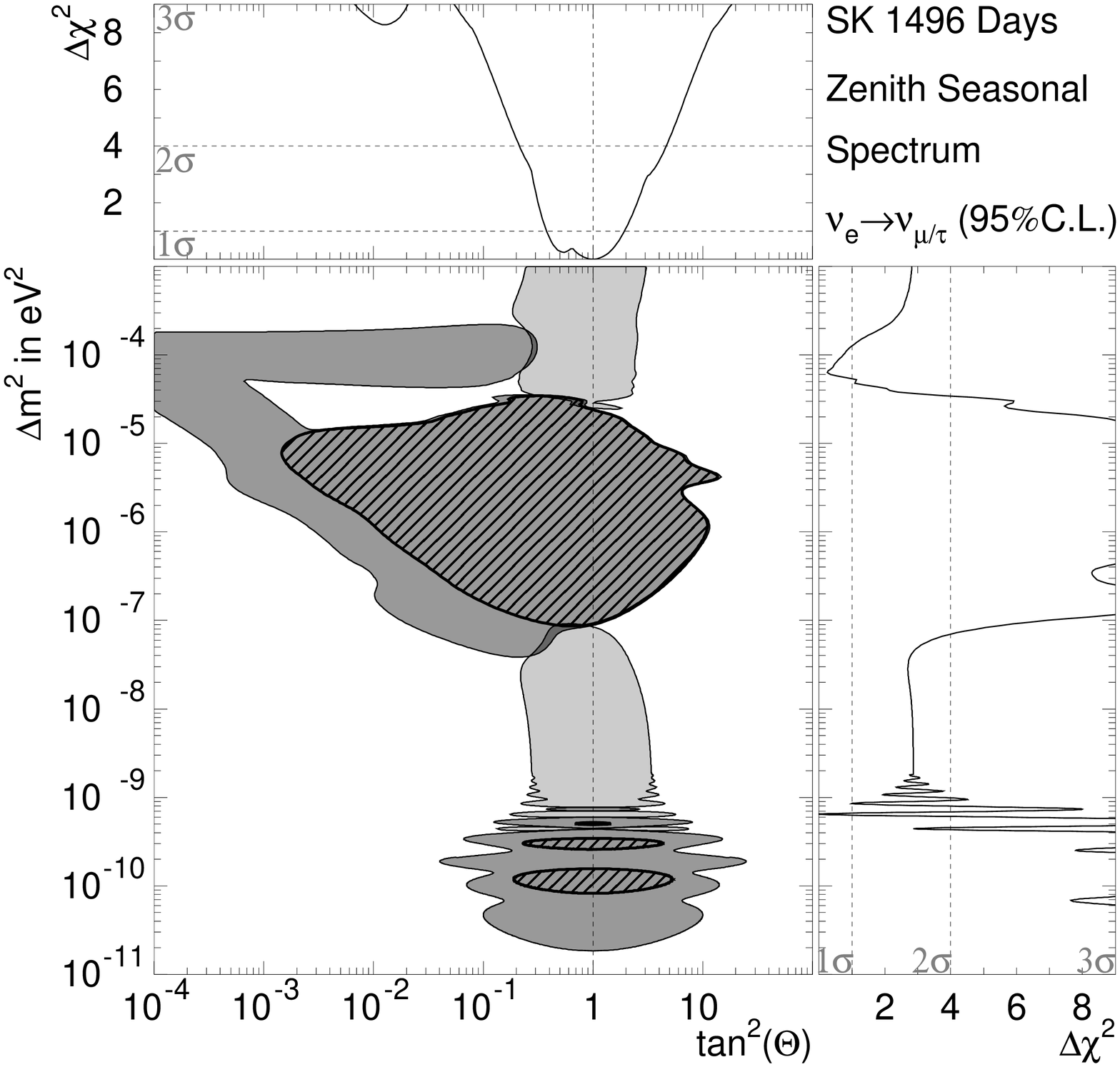}\hspace{0.2in}
\includegraphics[height=2.47in]{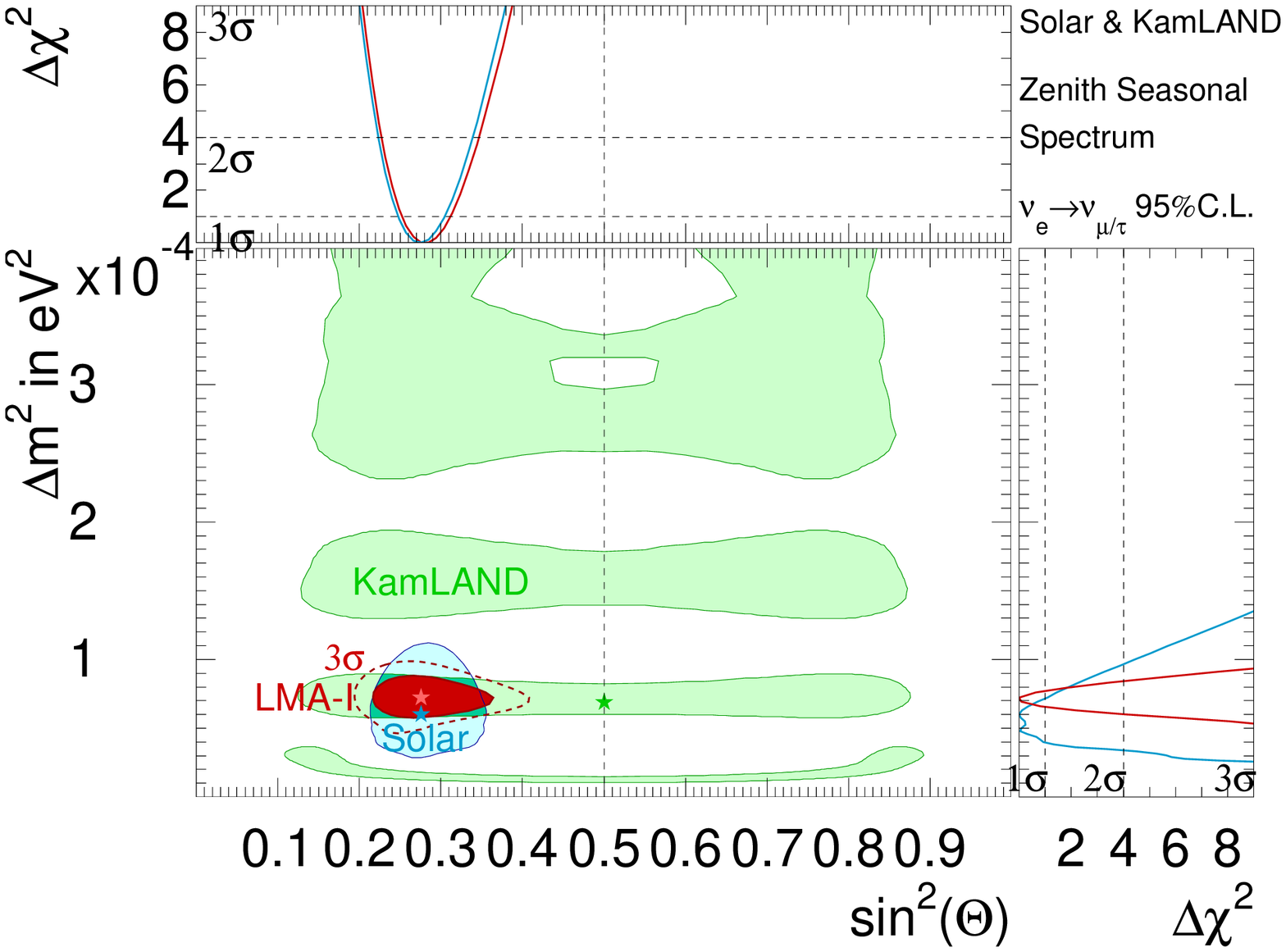}
\caption{Left: Excluded (SK spectrum and time variation; dark gray)
and allowed (SK spectrum, rate and time variation; light gray)
neutrino oscillation parameter regions at 95\,\% C.L.
Overlaid are the areas excluded by the day/night and seasonal variation
 (hatched regions).
The plots at the top (right) show the $\chi^2$ difference
as a function of $\tan^2\theta$ ($\Delta m^2$).
Right: Allowed region at 95\,\% C.L from all solar data, KamLAND results, and the combination of
all solar and KamLAND data. The plots at the top and right show the $\chi^2$ differences.}
\label{fig:skosc}
\end{center}
\end{figure*}

To constrain neutrino oscillation parameters using the time variations of solar neutrino flux,
the likelihood difference 
$\Delta\log {\cal L}=\log {\cal L}(\alpha=1)-\log {\cal L}(\alpha=0)$
between the expected time variation and no time variation is computed.
$\Delta\log {\cal L}$ is interpreted as a time-variation 
$\Delta\chi^2_{\mbox{tv}}=-2\Delta\log {\cal L}$ and is added to the spectrum $\chi^2$.
The total $\chi^2$ is defined as~:
\begin{equation}
\chi^2=\sum_{i=1}^{N_{\mbox{\tiny bin}}}\left(\frac{d_i-\rho_i}{\sigma_i}\right)^2
+\frac{\delta_B^2}{\sigma_B^2}+\frac{\delta_S^2}{\sigma_S^2}
+\frac{\delta_R^2}{\sigma_R^2}+\Delta\chi^2_{\mbox{tv}}
+\left(\frac{\beta-1}{\sigma_f}\right)^2
\end{equation}
where $d_i$ is the ratio of the data to the SSM prediction in the energy bin $i$ and $\rho_i$
 is the prediction in which neutrino oscillation is considered. $\sigma_i$ is energy bin-uncorrelated uncertainty.
$\sigma_B$, $\sigma_S$ and $\sigma_R$ are the estimated uncertainties in the $^8$B neutrino
spectrum, SK energy scale and SK energy resolution, respectively. The last term constraining
 the $^8$B flux to the SSM is optional.
The left-hand plot in Figure~\ref{fig:skosc} shows the allowed and excluded
 regions at 95\,\% C.L. obtained from the SK solar neutrino data.
The advantage of this method is that the number of energy bins can be increased because 
the analysis is independent of the binning in night time.

Stronger constraints on neutrino oscillation parameters result from the combination of the
 SK measurements with other solar neutrino data~\cite{solar} and the KamLAND results~\cite{kl}. 
The 95\,\% C.L. allowed region is shown in the right-hand plot in Figure~\ref{fig:skosc}.
 Only LMA solution remains. The allowed range of oscillation parameters
 are obtained to be 
$\Delta m^2$ = $7.2^{+0.6}_{-0.5} \times 10^{-5} eV^2$ and
$\tan^2\theta = 0.38\pm0.08$.

\subsection{Summary}
 Zenith angle analysis and L/E analysis were carried out using atmospheric
 neutrino data. 
 A dip in the $L/E$ distribution was observed for the first time, as predicted
 from the sinusoidal flavor transition probability of neutrino oscillation.
The allowed neutrino oscillation parameter region was constrained to be
 $1.9 \times 10^{-3}\,{\rm eV}^2 < \Delta m^2 
< 3.0 \times 10^{-3}\,{\rm eV}^2 $ and 
$\sin^{2}2\theta~ > 0.90$ at 90\,\% C.L.,
 which is consistent with that of the zenith angle analysis.

 Super-Kamiokande has precisely measured the solar neutrino flux,
recoil electron spectrum, and time variations of the flux.
No significant time variation and energy distortion appeared from the 1496 live
days of data. The time variation of neutrino flux was considered in neutrino
 oscillation analysis by a maximum likelihood method.
 Combined with all solar neutrino experiments and KamLAND data, 
 neutrino oscillation parameters are obtained to be
 $\Delta m^2$ = $7.2^{+0.6}_{-0.5} \times 10^{-5}
eV^2$ and $\tan^2\theta = 0.38\pm0.08$.

\end{document}